# The design optimization and experimental investigation of the 4.4 μm Raman laser based on hydrogen-filled revolver silica fiber

A. N. Kolyadin, M. S. Astapovich, A. V. Gladyshev, A. F. Kosolapov, A. D. Pryamikov, K. G. Alagashev, M. M. Khudyakov, M. E. Likhachev, I. A. Bufetov

*Fiber Optics Research Center, Russian Academy of Sciences, 38 Vavilov st., Moscow, 119333, Russia*

**Annotation.** Optical properties of hollow-core revolver fibers are numerically investigated depending on various parameters: the hollow-core diameter, the capillary wall thickness, the values of the minimum gap between the capillaries, the number of capillaries in the cladding and the type of glass (silica and chalcogenide). Preliminary, similar calculations are made for simple models of hollow-core fibers. Based on the obtained results, the optimal design of the revolver fiber for Raman laser frequency conversion (1.56 μm → 4.4 μm in $^1H_2$) was determined. As a result, efficient ns-pulsed 4.42 μm Raman laser based on $^1H_2$-filled revolver silica fiber is realized. Quantum efficiency as high as 36 % is achieved and output average power as high as 250 mW is demonstrated.

## Introduction

Silica hollow-core fibers (HCF) allow transmitting light in a much wider wavelength range [1] than solid-core silica fibers. In particular, silica HCFs can be used in the mid-IR spectral range (λ > 3 μm). As far as mid-IR range is concerned, one of the most promising HCF designs is the so-called revolver fibers (RF), the typical cross-sections of which are shown in Fig.1. When the light propagates through a revolver fiber, the fraction of optical power localized in the silica glass is as low as about $10^{-4}$. This circumstance makes it possible to achieve low-loss transmission in RF even in those spectral ranges where the silica glass has a strong fundamental absorption, for example, in the mid-IR range. At the same time, an important advantage of the revolver fibers is the simplicity of their design and fabrication as compared with hollow-core photonic-crystal fibers [2].

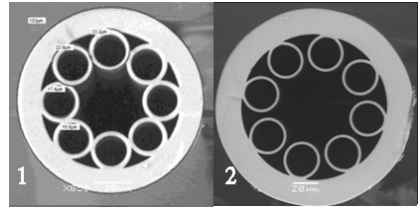

Fig.1. Hollow-core revolver fibers [3, 4].

Being filled by various gases, RF becomes a convenient active medium for mid-IR light generation. For example, by using silica RFs filled by molecular hydrogen isotopes, the Raman fiber lasers have been recently demonstrated at the wavelengths of 2.9, 3.3, 3.5, and 4.4 μm [5, 6].

To realize an efficient Raman fiber laser the optical properties of revolver fiber should be optimized for the Raman conversion of interest $\lambda_P \to \lambda_S$. A similar optimization problem was considered previously for the case of Raman lasers based on ordinary glass fibers [7]. It was shown that a single parameter $P_F$ can be introduced as a figure of merit (FOM) to evaluate an optical fiber as an active Raman medium for a single-stage Raman fiber laser:

$$P_F = \left( \sqrt{\frac{\alpha(\lambda_P)}{g_0}} + \sqrt{\frac{\alpha(\lambda_S)}{g_0}} \right)^2 \qquad (1)$$

The parameter $P_F$ takes into account simultaneously several optical fiber properties such as the optical losses at the pump $\alpha(\lambda_p)$ and Stokes $\alpha(\lambda_s)$ wavelengths, and the Raman gain of the fiber $g_0$ (e.g., in dB/(m*W)) for the Raman conversion $\lambda_P \rightarrow \lambda_S$. Although this figure of merit was first derived for solid-core fibers, the derivation procedure did not use the fact that the core was solid. Thus, the parameter $P_F$ can be equally applied to evaluate the gas-filled hollow-core fibers as a Raman medium. The physical meaning of the $P_F$ (having the dimension of Watts) is as follows [7]: it is the threshold pump power for a fiber Raman laser placed in a special high-Q resonator. Therefore, the smaller the $P_F$, the closer the fiber geometry is to the optimal design for the selected Raman conversion $\lambda_P \rightarrow \lambda_S$.

To optimize the RF design using expression (1), it is of high importance to determine the dependence of optical losses on the fiber parameters such as the hollow-core diameter, the capillary wall thickness, the value of the minimum gap between the capillaries, the number of capillaries in the cladding and so on. In this work we numerically investigate the optical losses of the revolver fibers as a function of various fiber geometry parameters in order to optimize the fiber design for 1.56 µm → 4.4 µm Raman conversion in a hydrogen-filled revolver fiber. Both direct numerical simulations and simplified analytical models are used to calculate the optical losses for various fiber designs. Then, a close-to-optimum revolver silica fiber is fabricated and used for experimental demonstration of an efficient 1.56 µm → 4.4 µm Raman conversion.

## RF design optimization based on analytical models.

A direct numerical simulation of the optical properties of the RF is a complex task, which requires relatively long calculation time. To speed up an optimization of the fiber geometry parameters, the simplified models of a hollow-core fiber can be used, especially those models that allow deriving analytical expressions for optical losses of the fiber. The waveguiding properties of the revolver fibers can be relatively simply explained by using a successive-approximation method to describe the propagation of the light along the fiber:

1. In the well-known model of a hollow circular waveguide (HCW) [8] the guiding structure is represented as a hollow dielectric pipe with thick walls made of glass. A silica glass (SG) and chalcogenide glass (CG) is of the greatest interest for us. In this model the waveguide properties are provided by the Fresnel reflection of radiation from the inner surface of the pipe in the case of grazing incidence (see Fig. 2a, b).
2. The reflection coefficient from the pipe wall can be increased in comparison with the HCW by using a thin pipe wall with reflection from two surfaces and a wall thickness corresponding to the constructive interference of the reflected rays. Such a model of a tubular optical fiber (TF) was considered in [9] (see Fig. 2c, d). In fact, the capillary wall in the TF model is a Fabry-Perot interferometer, which has the highest reflectivity when the anti-resonance condition is fulfilled. Much later, the same mechanism was considered in [10] and was called ARROW (Anti Resonant Reflecting Optical Waveguide).
3. Being added to the two previous mechanisms (Fresnel reflection and constructive interference of the light reflected from two surfaces), the negative curvature of the core-cladding boundary [3] reduces the optical losses even further by more than an order of magnitude (depending on other design parameters). This effect is a result of the increase in the reflection coefficient due to a decrease in the angle of incidence when the rays are reflected from the surface with negative curvature. In the presence of resonant energy transfer from the core to the cladding modes a corresponding local change in the optical losses can be observed. This is a more subtle effect, which can no longer be explained in the ray approximation.

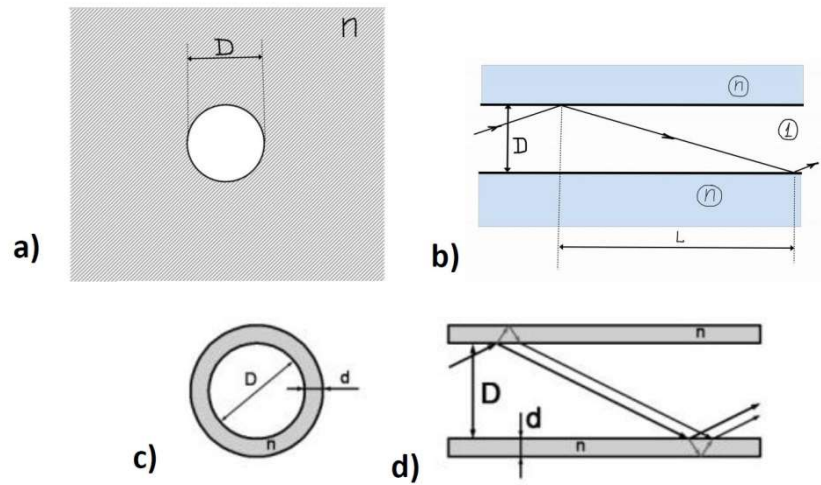

Fig. 2. a) and b) - hollow circular wavegiude (HCW) model - hollow dielectric pipe with thick walls; a)-the cross-section and b) scheme of the radiation propagation along the HCW axis; c) and d) tubular optical fiber (TF) model - a capillary with walls thin enough to reveal interference phenomena in the light reflection ; c) TF cross-section, d) radiation propagation scheme along the TF axis.

Among mentioned above approximations the first two (HCW and TF) provide analytical solutions for optical losses. Both models can be considered as the predecessors of revolver fibers. The HCW model overestimates the value of optical losses because of a lowered reflection coefficient of radiation at the core-cladding interface, since radiation is reflected from only one interface. The TF model leads to smaller optical losses, since the reflection coefficient at the core-cladding interface can be increased because a thin capillary wall acts as a reflective Fabry-Perot filter. But it must be taken into account that both of these analytical models do not take into account the effect of the negative curvature of the core-cladding interface in the RF, which both theoretically and experimentally leads to an additional (and significant) decrease in optical losses in the RF in comparison with both the TF and especially the HCW models. Nevertheless, the HCW and TF models can apparently be used to determine a number of features of the revolver fibers prior to starting a large-scale numerical simulation of real RF.

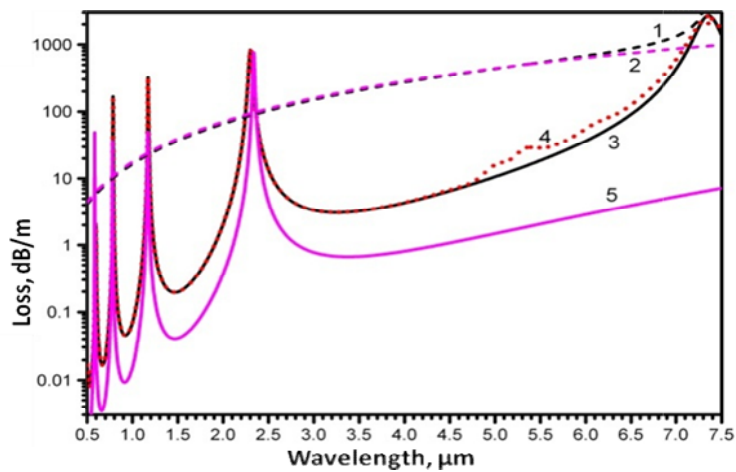

Fig. 3. Spectral dependencies of optical losses for various models of straight HCFs. For all models, the hollow core diameter is assumed to be 77 μm. 1- losses in silica HCW. 2-optical losses in chalcogenide HCW. 3- losses in silica TF without taking into account material losses; 4- TF with taking into account material losses; 5- losses in the chalcogenide TF.

As an example, Fig. 3 (line 1) shows the calculated optical loss spectrum for the fundamental mode of a silica HCW with a hollow core diameter of 77 μm (hereinafter only losses for the fundamental mode of the fiber are considered). The losses in such an optical fiber in the mid-IR range exceed ~100 dB/m.

Note, the optical losses have a maximum (like singularity) near the wavelength of 7.3 µm, where the real part of the refractive index of silica glass crosses the value Re($n_{SiO2}$)= 1. In this case, the reflection at the glass-air interface practically disappears, which leads to a sharp increase in losses. The waveguiding effect is virtually absent, and this is true for any HCF design made of silica glass. In the case of a RF made of chalcogenide glass (refractive index ≈2.5 in the middle IR range, see [11], Table 5.4), there is no singularity in optical loss spectrum at 7.3 µm is (Fig. 3, line 2).

The optical loss spectrum of the TF (in contrast to the HCW) has the form of a sequence of transmission bands. The bands appear due to the resonant properties of the thin wall of the capillaries that form the core-cladding interface in the TF model (Figure 3, line 3). The boundaries of the bands are determined by the condition of resonant transmission of the glass cylindrical film (see [10]). The minima of optical losses are reached approximately at the center of the spectral bands, and the optical loss values at these wavelengths are much lower than in the HCW. The longest wavelength transparency zone ("zero" band) of the TF has a number of features due to a significant and nonmonotonic variations in the complex refractive index of silica glass in the mid-IR range. Here, as in the case of HCW, a peak of optical losses near 7.3 µm is observed. The peak is explained solely by the fact that the real part of the refractive index in this region is close to 1. Therefore, optical fibers made of silica glass with a hollow core lose their waveguiding properties in the vicinity of λ≈7.3 µm. This fact must be taken into account for correct interpretation of the optical phenomena observed in such fibers [12].

Absorption in the silica glass at wavelengths of about 4.5 µm already reaches ~ 3000 dB/m, and in the range 4.5-5 µm there is a significant further increase in optical losses to ~ 50000 dB/m [1]. Such a strong material absorption gives an additional contribution to the total losses of the TF. For example, Fig.3 illustrates that at wavelengths above ~4.7 µm the total optical losses in the TF (Fig. 3, line 4) start to differ markedly from the waveguide losses of the fiber (Fig. 3, line 3). Finally, the total optical losses of the chalcogenide TF (Fig. 3, line 5) are about an order of magnitude lower compared with losses of a silica fiber and does not have opacity peak in the vicinity of 7.3 µm wavelength.

When the thickness of the capillary wall is changed, the TF model predicts the spectral shift of the transmission bands. As a result, the minimum values of the optical losses in the TF are reached at different wavelengths. It makes sense to consider the dependence of the value of the minimally attainable optical losses in such a fiber for each wavelength (with an appropriate variation of the thickness *d* of the capillary wall (see Figs 2, c and d) [9]. Figure 4 shows such a dependence for silica glass without taking into account the absorption of silica glass (line 1) and taking into account the absorption in the silica glass, provided that the wavelength corresponds to the center of the "zero" (line 2) and "second" (line 3) transmission band. The comparison of these lines shows that material losses can be neglected at short wavelengths only, but at wavelengths above ~4.7 µm they increase the total optical losses to the values above 1 dB/m (as was noted in [13].) Figure 4 also shows the analogous dependence for RF made of chalcogenide glass (line 5). This curve was obtained for the same conditions as were used to calculate the line 2 in the case of silica glass. Finally, for comparison, the calculated loss spectra for the HCW model for the CG (line 4) and the SG (line 6) are also given in the Figure 4. It should be noted that the TF model [9] uses to some extent the planar fiber approximation, which leads to a dependence of the optical losses on the polarization of the radiation. In a circular cylindrical TF, such dependence, of course, is absent. In this case, we used data for polarization with the lowest attenuation. For a real fiber, the losses depend on the wavelength in qualitatively the same way, although the absolute value of losses can be significantly higher.

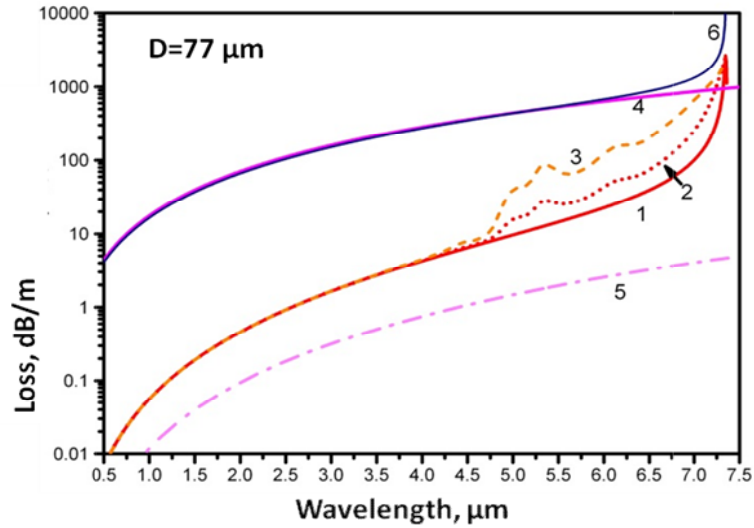

Fig. 4. Spectra of minimum values of optical losses for TF (lines 1-3 and 5) and HCW (lines 4 and 6). For all lines, the hollow core diameter is assumed to be 77 μm. 1-spectrum of losses of silica TF without regard for material absorption, 2 and 3, taking into account material absorption (the thickness of the capillary wall corresponds to the middle of the zero (2) and second (3) transparency zones), 5-spectrum of losses of the chalcogenide TF. For comparison, the loss spectra of chalcogenide HCW (4) and silica HCW (6) are shown.

Thus, the above estimates show that hollow-core optical fibers made of silica glass can have optical losses of ~1 dB/m at wavelengths below ~4.7 μm. (We emphasize once again that for real revolver fibers the values of the optical losses shown in Figures 3 and 4 should be considered as the upper limit estimates). In the case of chalcogenide HCF, the material absorption in the mid-IR range is small enough to have practically no effect on the total losses in the HCF. As can be seen in Fig. 3 and Fig. 4, there is no additional maximum at ~7.3 μm in the optical loss spectra of chalcogenide HCF. Compared with losses of a silica HCF, the total loss level of chalcogenide HCF is about an order of magnitude lower due to the higher reflectivity of each air-glass interface. As one would expect, chalcogenide fibers allow in principle to obtain HCF with substantially lower losses in the mid-IR region. But chalcogenide glasses still face many technological challenges and a satisfactory technology for HCF fabrication has not been yet developed [11].

It is known that for straight (i.e., non-bent) HCW and TF the optical losses depend on the diameter of the hollow core as $1/D^3$ and $1/D^4$, respectively [14]. Therefore, the both HCW and TF models predict that for a straight fiber the figure of merit $P_F$ is proportional to $1/D$ (for the HCW) and $1/D^2$ (for the TF). Consequently, the larger D, the smaller the $P_F$. Thus, there is no optimum for the straight optical fibers with respect to the diameter of the hollow core. However, bent-induced losses must be also considered. The ability to bending is one of the main advantages of optical fibers. Assuming that we are working with HCW and TF optical fibers that are coiled with a certain radius R, the bent-induced losses in such fibers grow as the diameter of the hollow core D grows. As a result, the optical loss of such fibers (and the figure of merit $P_F$) has a minimum at some value of the hollow core diameter $D_{min}$, which determines the optimal diameter of the hollow core of the fiber to a given hydrogen pressure in the core. Fig. 5 shows the spectra of optical losses of HCW and TF made of silica and chalcogenide glasses for wavelengths of 1.56 μm and 4.4 μm and a bend radius of R = 15 cm.

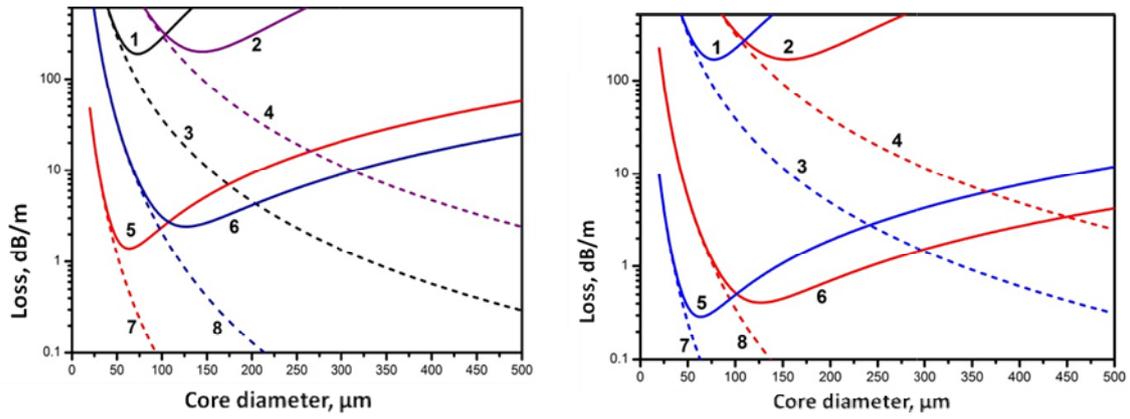

Fig. 5. Variation of optical losses for HCW and TF made of silica glass (a) and chalcogenide glass (b) wound on a coil with a radius of 15 cm with hollow core diameter. a) and b): HCW loss at a wavelength λ= 1.56 μm (line 1) and at λ = 4.4 μm (line 2); the spectra of optical losses for TF at a wavelength of λ = 1.56 μm (line 5) and at λ = 4.4 μm (line 6). Dashed lines (lines 3,4,7,8) show everywhere the course of optical losses without taking into account the bending of the fiber.

Minimal losses in optical fibers are achieved at different diameters of the hollow core for different wavelengths. Thus, in the case of HCW, the minimal losses correspond to the core diameter of $D_{min}$(1.56 μm) = 73 μm and $D_{min}$(4.4 μm) = 145 μm for silica fibers and to $D_{min}$(1.56 μm) = 78 μm and $D_{min}$(4.4 μm) = 155 μm for chalcogenide fibers. In the case of TF, $D_{min}$(1.56 μm) = 63 μm and $D_{min}$(4.4 μm) = 129 μm for silica fibers, while $D_{min}$(1.56 μm) = 64 μm and $D_{min}$(4.4 μm) = 126 μm for chalcogenide fibers. To determine the optimum core diameter for 1.56 μm→4.4 μm Raman conversion, we calculated the $P_F$ value according to expression (1) as a function of D. Fig. 6 shows the dependences $P_F$ (D) for silica (a) and chalcogenide (b) hollow-core fibers, which were modeled as a HCW and TF with a bent radius of R = 15 cm. It was assumed that the hollow core was filled by molecular hydrogen ($^1H_2$) under a pressure of 30 atm. The Raman gain in $^1H_2$ was assumed to be 1.2 cm/GW [15]. The minimum value of $P_F$ in a silica TF is achieved at D≈75 μm and is 700 W. In the chalcogenide TF, the minimum is reached approximately at the same diameter and is 126 W. In HCW minima are also attained at 75 μm with an accuracy of 1 μm, but the $P_F$ values at the minimum for this case reach $10^5$ W.

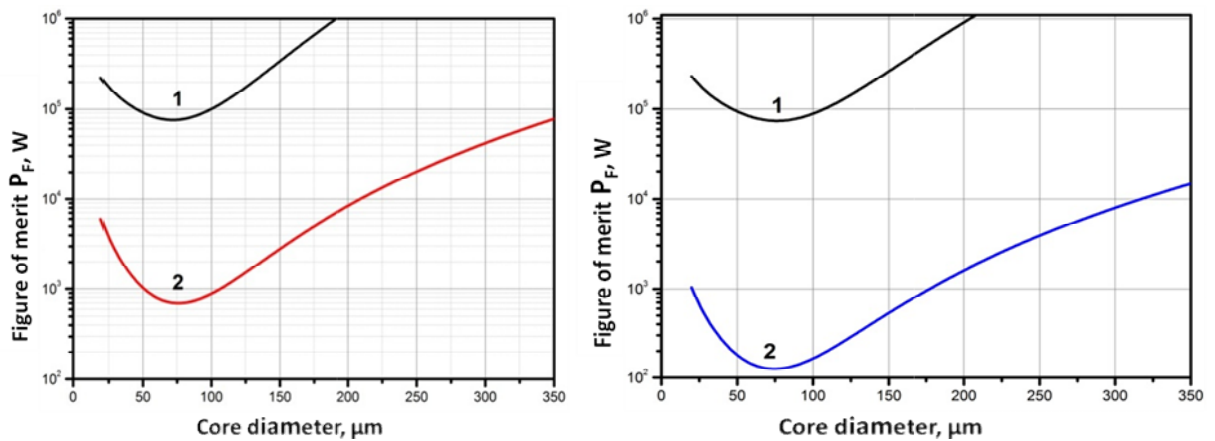

Fig. 6. Variation of the figure of merit (in watts) with the hollow core diameter for Raman fibers (according to HCW model (line 1) and the TF model (line 2)) filled with hydrogen at a pressure of 30 atm. a) the dependences for the fibers made of silica, b) - for fibers made of chalcogenide glass.

Thus, in an optical fiber of the HCW-type the threshold pump power of a 4.4 μm Raman laser is not less than 75 kW. Such a high value of $P_F$ is due to a sharp increase of optical losses in the HCW during bending. In fibers of the TF type, a $P_F$ minimum is achieved at approximately the same core diameter, but it is about 700 W for silica glass fibers. The usage of chalcogenide glass will reduce the $P_F$ value by a factor of ~6. Note, that despite the optical losses of HCW and TF differ by a factor of ~ 100 the optimal core diameters are fairly close in value. Also, the difference in optical properties of silica and chalcogenide glasses has only little effect on the value of optimal diameter of the hollow core.

Revolver fibers, in comparison with TFs, have one more mechanism that reduces the optical losses. This mechanism originates from the negative curvature of the core-cladding boundary. As a result, the losses are further reduced by more than order of magnitude. Nevertheless, the optimal diameter of the hollow core should be obviously expected in the vicinity of D = 75 µm.

## RF design optimization based on numerical simulations.

For a comparative analysis of the revolver fibers for a 1.56 µm→4.4 µm Raman hydrogen laser (RHL), we performed a numerical simulation of the optical properties of RFs, depending on the values of their geometric parameters. As indicated above, and as previous studies showed [5, 6], the revolver fibers are promising for the development of efficient Raman fiber gas lasers in the mid-IR range. In the process of optimization, the following RF's parameters were considered: the number of capillaries in the cladding *N*, the capillary wall thickness *d*, the minimum distance between the neighboring capillaries *t*, the diameter of the hollow core *D* (the maximum diameter of the circle inscribed in the core of the RF), and the refractive index of the glass *n*. To find the optimal RF's design for RHL, we have modeled several types of RFs with a cladding consisting of a single layer of capillaries made of pure silica glass and chalcogenide glass with a chemical composition of $As_2S_3$.

An analysis of the optical properties of silica RFs was carried out for the case of *N* = 10, 9, 8, and 7 capillaries in a cladding. The real part of the refractive index of the SG at a wavelength of 4.4 µm is *n* = 1.372. The material losses of the SG at this wavelength are 3372 dB/m. The numerical simulation was carried out in such a way that the wavelength of 4.4 µm should be in the zero transmission band, while the pump wavelength of 1.56 µm was in the first transmission band. The value of the capillary wall thickness *d* determines the width and position of the transmission bands of the RF. The process of optimizing the thickness of the capillary wall *d* was to ensure that the selected thickness corresponded to the minimum value of the loss of the RF at the Stokes wavelength (in the zero transmission band). At the same time, it was assumed that the selected thickness of the capillary wall *d* not necessarily put the pump wavelength in the minimum of the first transmission band. This is justified by the fact that the waveguide losses of the silica RF in the wavelength region 1.56 µm are very small, as, indeed, the material losses of the SG.

The analysis showed that the most optimal structure for the implementation of the RHL is the structure of a RF with N = 7: a 7-capillary silica revolver fiber. In order to verify this, calculations were carried out for losses in the zero and first transmission bands, depending on the thickness of the capillary wall (the calculation was carried out only in the region of antiresonant reflection of the capillary wall) at three values of the distance between them 0.1, 1, and 5 µm. The loss curves for a RF with N = 10, 9, and 8 capillaries in a cladding had a resonant character, i.e. inside the transmission bands there are additional local increases in optical losses, caused by resonant energy transfer between the fundamental mode of the hollow core and the cladding modes [16]. Only at a relatively large distance between the capillaries *t* = 5 µm these resonances are not observed. But at the same time, because of the increase in *t*, the waveguide losses in the RFs were increased due to the penetration of radiation from the modes of the hollow core into the space between the capillaries. The loss curves of a RF with 7 capillaries in a cladding have a resonance-free character both at small distances between the capillaries and at large distances. This allows us to have technological advantages in the form of additional tolerances in the manufacture of such RFs. In addition, with N=7, the losses increase insignificantly as the distance between the capillaries increases.

A similar analysis was performed for chalcogenide RFs. In this case, the wavelength of the Stokes wave also occurred at the zero transmission band, while the pump wavelength was placed in the 1st or 2nd transmission band, depending on the thickness of the capillary wall. It is clear that since the real part of the refractive index of the considered chalcogenide glass is equal to 2.41 (here the imaginary part is small and corresponds to the optical losses of about several dB/m at a wavelength of 4.4 µm), in order to obtain the position of the Stokes wave in the zero band the capillary wall thickness must be lowered (compared with the silica RF). Such a reduction in the thickness of the capillary wall inevitably must lead to additional difficulties in the process of their manufacture. Therefore, we also considered a variant with a pump wavelength in the second transparency band.

As in the case of silica RFs, we optimized the geometric parameters of the chalcogenide RFs and showed that the most suitable fiber for RHL in this case is also the 7-capillary fiber, since it has resonance-free loss curves, as for small (0.1 µm) and at large (5 µm) distances between the capillaries. Values of $t > 5$ µm were not considered, since a further increase in $t$ leads to an increase in the waveguide losses.

Here we consider dependences of losses for silica and chalcogenide RFs with a cladding consisting of seven capillaries depending on diameter of a hollow core. From technological point of view, the most optimal is the RF with the largest distance between the capillaries (in order to prevent them from sticking together during drawing), so in calculation the distance between the capillaries of the cladding was 5 µm, the wall thickness of the capillary was 1.29 µm (for silica RF). In the case of chalcogenide RF, the distances between the capillaries were also 5 µm, while the wall thickness was 0.6 µm. Fig. 7 shows the dependence of losses for the RF with seven capillaries in the cladding, depending on the diameter of the hollow core. When the diameter of the core was changed, the thickness of the walls of the capillary was kept constant (to fix the position of the transmission bands), and the diameter of the capillaries changed in such a way that the gap between the capillaries $d$ remained constant. In addition, it was assumed that the fiber axis is a straight line (i.e., the fiber does not have a bend).

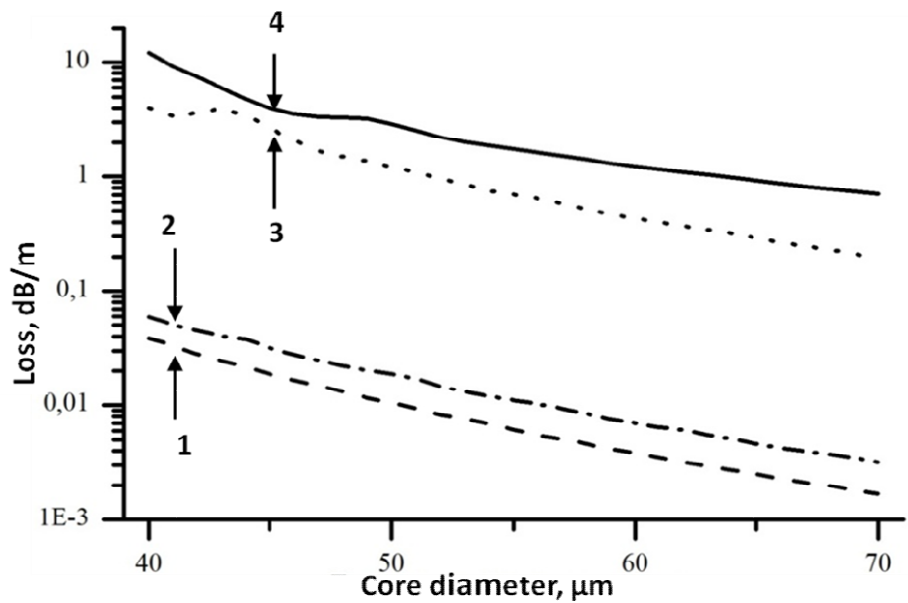

**Fig.7. Variation of the total losses of the single mode of the hollow core for silica (lines 1 and 4) and chalcogenide RF (lines 2 and 3), having 7 capillaries in the cladding, with the diameter of the hollow core at a pump wavelength of 1.56 µm (lines 1 and 2) and at a Stokes wavelength of 4.4 µm (lines 3 and 4). The distance between the capillary walls is 5 µm.**

Fig. 7 shows that the loss of chalcogenide RF at a wavelength of 4.4 μm is lower than in the case of silica RF, and this difference increases with the core diameter. A reverse situation is observed at the pump wavelength. Such a small difference in the magnitude of the losses at a wavelength of 4.4 μm with such a large difference in the material losses of the SG and CG is explained by comparable waveguide losses in the silica and chalcogenide RF. In addition, the difference in losses at a wavelength of 1.56 μm is also due to the fact that in both cases the parameters of the RF's structure were chosen on the basis that at a wavelength of 4.4 μm (zero transmission band) there should be minimal losses. Therefore, the pump wavelength of 1.56 μm (in the next transmission band) could not coincide with the minimum of the loss in the corresponding transmission band.

Thus, for further optimization the RF design with seven capillaries in the cladding was chosen, the distance between them being 5 μm, and the thickness of the capillary wall ensured that the pump wavelength and Stokes wavelength are in the region of low losses of the corresponding optical transparency band. When the diameter of the core was varied, the diameter of the capillaries in the reflecting cladding was also necessarily changed. With increasing $D$, for example, from 40 to 70 μm, the outer diameter of each capillary increased from 14 to 45 μm. Optimization of the diameter of the hollow core was carried out as described below.

As follows from the preliminary consideration (see above), for a straight HCF as an active element of a Raman gas laser there is no optimal diameter of the hollow core: theoretically, the larger the diameter, the better. In reality, the use of a straight fiber is technically difficult, and when creating a RF with large $D$, there are great technological difficulties. If the fiber is used in the usual way, i.e. coiled with some radius $R$, then the situation with the optimal core diameter changes significantly. Assuming that the optical fiber is coiled with a diameter of 30 cm ($R$ = 15 cm), the values of the optical losses were calculated for the RF at pump wavelengths (1.56 μm) and the Stokes component (4.4 μm, see Fig. 8). Also we calculated the Raman gain coefficient for RF filled with hydrogen under a pressure of 30 atm. Obtained data were used to calculate the figure of merit of real RF as an active Raman medium.

Fig. 8 shows the dependence of the optical losses in the RF on the diameter of the core.

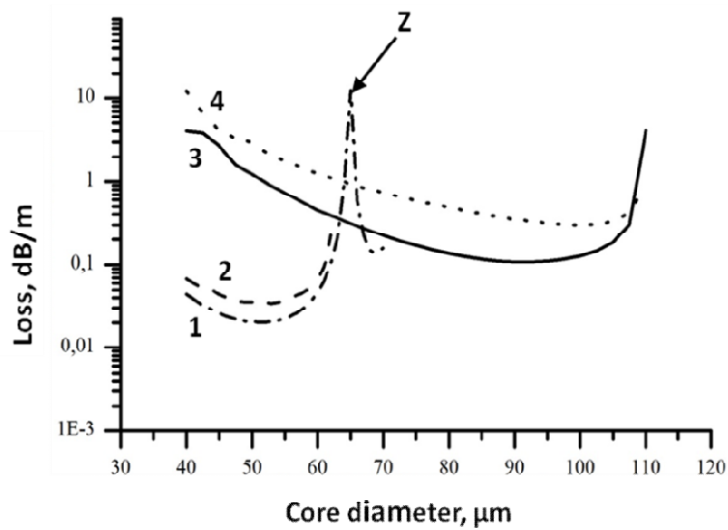

**Fig. 8. Loss of the fundamental mode for silica (lines 1 and 4) and chalcogenide (lines 2 and 3) 7-capillaries RFs vs hollow core diameter at the wavelength 1.56 μm (lines 1 and 2) and 4.4 μm (lines 3 and 4). The RFs are bent with a radius of 15 cm. Z is the resonance peak due to the fiber bending.**

The minimum value of optical losses at a wavelength of 4.4 μm for both types of RFs (SG and CG) is achieved approximately at D = 100 μm, while the loss of RF from the CG is several times lower than the loss in the silica RF. The minimum value of the optical losses at the pump wavelength corresponds to 50 μm of the hollow core diameter in both cases. It should be noted that the values of *D*, at which minimum losses are attained both at 4.4 μm and 1.56 μm, are in good agreement with the results of the TF model. This suggests that the TF model describes quite well some of the waveguide properties of a RF with a large hollow core diameter. The overall level of RF losses is much lower than in the case of HCW and TF, the introduction of the negative curvature of the core-cladding boundary affects here. The resonant peak observed in the dependence of the losses at the pump wavelength on *D* (Fig. 8) is associated with the excitation of the capillary mode [16].

Now, substituting the obtained data into expression (1) for the figure of merit of the RF, we obtain the $P_F(D)$ dependence for the real fiber, presented in Fig. 9.

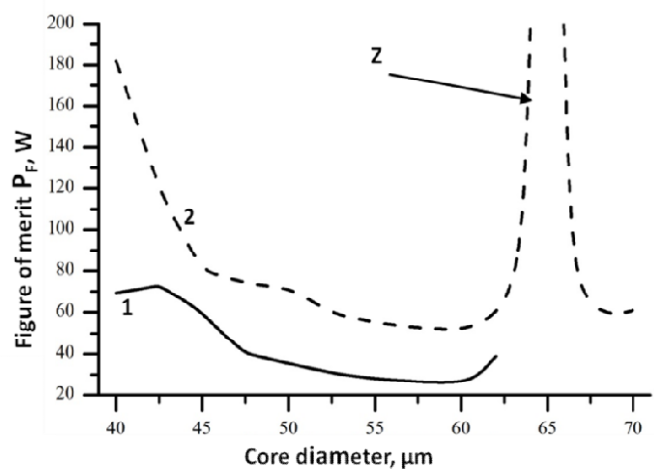

Fig. 9. Variation of the FOM $P_F$ with the diameter of the hollow core for the fundamental mode of 7-capillary silica (line 1) and chalcogenide (line 2) RF filled with hydrogen at 30 atm. Z is the resonance peak due to the fiber bending.

As can be seen from Fig. 9, the value of the figure of merit for silica and chalcogenide RF reaches a minimum in the region of values of the hollow core diameter of 60 μm, which roughly coincides with the analogous value for TF and HCW. At the same time, the value of the figure of merit for silica RFs is approximately three orders of magnitude lower than in the case of HCW and 15 times smaller than in the case of TF. For a chalcogenide RF the difference in the figure of merit compared to a HCW is approximately the same, and in the case of a TF, it is approximately 2 times smaller. The resonant peak in the dependences of the figure of merit value is related to the same bent-induced excitation of the capillary mode, which was described above.

### Experimental demonstration.

No doubt, the above estimations indicate that compared to silica HCFs the chalcogenide HCFs are preferable for the 4.4 μm laser development. But chalcogenide glass is much less manufacturable than silica. While silica RFs have minimal value of $P_F$ about 60 W, that is quite reasonable value for the peak powers of nanosecond pulsed fiber pump lasers at 1.56 μm. So, for technological reasons silica was chosen as a material for the RF fabrication.

For experimental investigation a revolver-type HCF with optimal core diameter of 75 μm was fabricated. The calculated fundamental mode field diameter was 55 μm and the cladding was formed by ten silica silica (F300) capillaries with 21.7 μm inner diameter, 1.15 μm wall thickness, and 6 μm the minimum distance between the neighboring capillaries. We increased the number of capillaries to ten with corresponding reduction in their diameters to avoid the appearance of resonance peaks like peak *Z* in Fig. 8 and Fig.9.

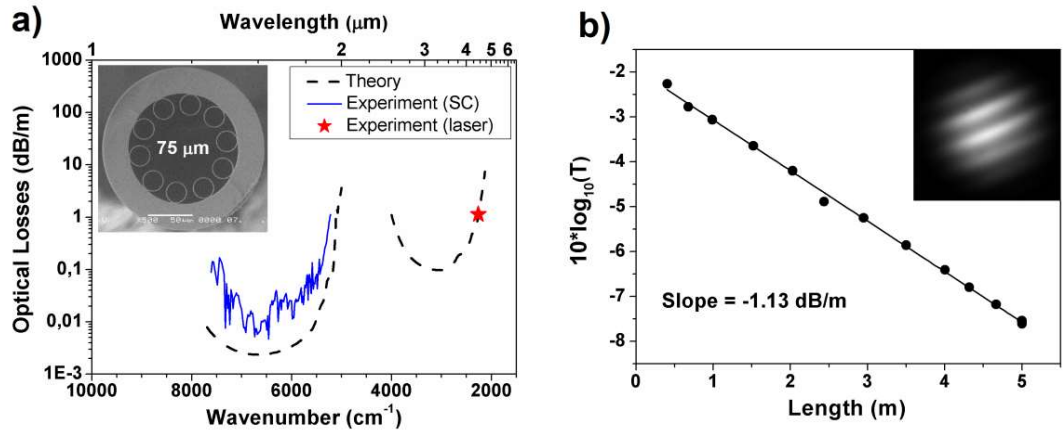

Fig. 10. (a) The spectrum of optical losses calculated for the revolver fiber (black dashed curves). The losses measured using supercontinuum source (blue solid curve) or using single-mode 4.42 μm laser (red star) are also shown. SEM image of the revolver fiber cross section is shown in the inset. (b) RF transmission measured at λ = 4.42 μm as a function of the revolver fiber length. The mode intensity distribution (inset) corresponds to a fundamental mode modulated by interference at the entrance window of the beam profiling camera.

The calculated loss spectrum of the fiber is shown in Fig. 10a (dashed curves). Measured optical losses (Fig. 10a, blue solid curve, red star) correlate well with the results of calculations. To analyze Raman conversion we are particularly interested in loss values for the pump (λ = 1.56 μm) and the Stokes (λ = 4.42 μm) waves. At the pump wavelength the measured attenuation has a value of 0.03 dB/m, which in fact is equal to the accuracy of the measurements. To determine optical losses at Stokes wavelength, fiber transmission (*T*) at this wavelength was measured as a function of the fiber length. As a result, optical losses at λ = 4.42 μm was found to be as low as 1.13 dB/m. Note, the material absorption of silica cladding at 4.42 μm wavelength amounts to ~ 4000 dB/m. The scheme of the experiment is described in detail in [17].

The output power measured for each spectral component is shown in Fig. 11 as a function of average pump power coupled to the fiber. In spite a large quantum defect for 1.56 → 4.42 μm conversion, the average power as high as 250 mW was generated at λ = 4.42 μm when coupled pump power was 1.9 W (Fig. 11, red dots). In this regime the Stokes pulses had duration of ~2 ns and peak power of ~2.5 kW. The power conversion efficiency of 13 % was achieved, corresponding to quantum conversion efficiency as high as 36 %.

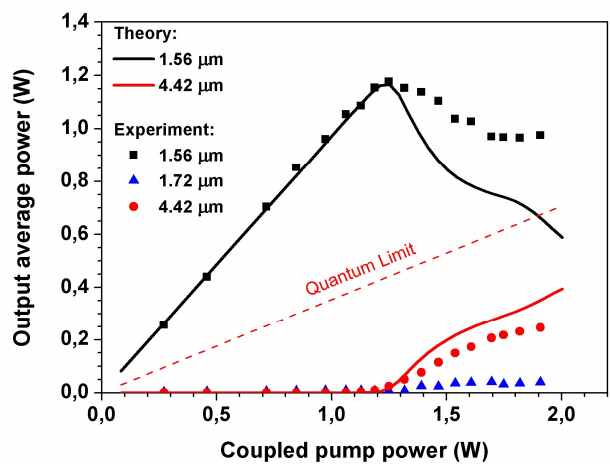

Fig.11. The average output power of the Raman laser as a function of coupled average pump power. Measured data are shown for residual pump (black squares), vibrational Stokes at 4.42 μm (red dots) and rotational Stokes at 1.72 μm (blue triangles). Results of numerical calculations are also shown for vibrational Stokes (red solid curve) and residual pump (black solid curve). Red dashed line represents a quantum limit for generation at 4.42 μm.

The measured (Fig. 11, dots) and calculated (Fig. 11, lines) power dependencies are in good agreement with each other. Some deviations of the measured power at λ = 4.42 μm from calculated values can be attributed to excitation of higher-order modes at the pump wavelength and, partially, to occurrence of rotational Stokes wave at λ = 1.72 μm.

## Conclusions.

Numerical analysis was carried out both for simplified models of a hollow-core fiber and for the model of a real revolver hollow-core fiber. The analysis made it possible to determine the optimal parameters of the structure of the hollow-core fiber for the hydrogen Raman fiber laser 1.56→4.4 μm. The obtained parameters were used as initial data for development a hollow-core fiber for the Raman laser. The manufactured fiber with parameters close to optimal ones allowed us to obtain high efficiency and low pump power threshold.

Some complication of the problem in the case of both types of RFs is the uprising of resonances in the dependence of optical losses on bending. But at the stage of RF's parameters calculation it is always possible to avoid such resonances by changing the parameters of the structure.

We experimentally demonstrated [17] the realization of RF with geometry close to optimal for RHL. The assembled RHL generates 250 mW at a wavelength of 4.4 μm with quantum efficiency of 36%.

**Funding**

Russian Science Foundation (RSF) (№16-19-10513).

**References and links**

1. A.D. Pryamikov, A.F. Kosolapov, G.K. Alagashev, A.N. Kolyadin, V.V. Vel'miskin, A.S. Biriukov, I.A. Bufetov. "Hollow-core microstructured 'revolver' fibre for the UV spectral range", Quantum Electronics, 46(12), 1129-1133 (2016).

2. Gregan R F, Mangan B J, Knight J C, Birks T A, Russell P St J, Roberts P J and Allan D C "Single mode photonic band gap guidance of light in air", Science 285б 1537-39 (1999).

3. Pryamikov A.D., Biriukov A.S., Kosolapov A.F., Plotnichenko V.G., Semjonov S.L., Dianov E.M. Opt. Express, 19 (2), 1441 (2011).

4. Kolyadin A.N., Kosolapov A.F., Pryamikov A.D., Biriukov A.S., Plotnichenko V.G., and Dianov E.M. "Light transmission in negative curvature hollow core fiber in extremely high material loss region" Optics Express, Vol. 21, Issue 8, pp. 9514-9519 (2013)

5. A.V. Gladyshev, A. N. Kolyadin, A. F. Kosolapov, Yu. P. Yatsenko, A. D. Pryamikov, A. S. Biriukov, I. A. Bufetov, and E. M. Dianov, "Efficient 1.9 μm Raman generation in a hydrogen-filled hollow-core fiber", Quantum Electron., vol. 45, no. 9, pp. 807–812 (2015).

6. A.V. Gladyshev, A. F. Kosolapov, M. M. Khudyakov, Yu. P. Yatsenko, A. K. Senatorov, A. N. Kolyadin, A. A. Krylov, V. G. Plotnichenko, M. E. Likhachev, A. Bufetov, and E. M. Dianov, "Raman Generation in 2.9 - 3.5 μm Spectral Range in Revolver Hollow-Core Silica Fiber Filled by H2/D2 Mixture," in CLEO, OSA Technical Digest (online) (Optical Society of America, 2017), paper STu1K.2

7. I. A. Bufetov and E. M. Dianov, "A simple analytic model of a CW multicascade fibre Raman laser," Quantum Electron., 30(10), 873–877 (2000).

8. E.A.J. Marcatili, R.A. Schmeltzer, "Hollow metallic and dielectric waveguides for long distance optical transmission and lasers", Bell Syst. Tech. J., 43, 1783-1809 (1964).

9. M.Miyagi. "Bending Losses in hollow and dielectric tube leaky wavegiudes", Applied Optics, 20(7), 1221-1229 (1981).

10. Litchinitser N.M., Abeeluck A.K., Headley C., Eggelton B.J. "Antiresonant reflecting photonic crystal optical waveguides" Opt. Lett., 27 (18), 1592 (2002).

11. Chalgenide glasses. Preparation, properties and applications. Edited by J.-L. Adam and X.Zhang. Woodhead publishing, Oxford, Cambridge, Philadelphia, New Delhi, 714 pp, (2014).

12. A. Benoît, B. Beaudou, M. Alharbi, B. Debord, F. Gerome, F. Salin, F. Benabid. "Over-five octaves wide Raman combs in high-power picosecond-laser pumped $H_2$-filled inhibited coupling Kagome fiber", Optics Express, 23(11), 14002-14009 (2015).

13. R. R. GATTASS, D. RHONEHOUSE, D. GIBSON, C. C.MCCLAIN, R. THAPA, V. Q. NGUYEN, S. S. BAYYA, R. J. WEIBLEN, C. R. MENYUK, L. B. SHAW, J. S. SANGHERA. " Infrared glass-based negative-curvature antiresonant fibers fabricated through extrusion", Optics Express 24(22) 25697-25703 (2016)

14. A. M. Zheltikov, "Colors of thin films, antiresonance phenomena in optical systems, and the limiting loss of modes in hollow optical waveguides" Phys. Usp. 51 591–600 (2008).

15. A. V. Gladyshev, A. N. Kolyadin, A. F. Kosolapov, Yu P. Yatsenko, A. D. Pryamikov, A. S. Biriukov, I. A. Bufetov, E. M. Dianov, "Low-threshold 1.9 mu m Raman generation in microstructured hydrogen-filled hollow-core revolver fibre with nested capillaries LASER PHYSICS", 27(2), 025101 (2017)

16. G. K. Alagashev, A. D. Pryamikov, A. F. Kosolapov, A. N. Kolyadin, A. Yu. Lukovkin, and A. S. Biriukov, Impact of geometrical parameters on the optical properties of negative curvature hollow core fibers, Laser Physics, vol. 25, 055101 (2015)

17. V. Gladyshev, A. F. Kosolapov, M. M. Khudyakov, Yu. P. Yatsenko, A. N. Kolyadin, A. A. Krylov, A. D. Pryamikov, A. S. Biriukov, M. E. Likhachev, I.A. Bufetov, E. M. Dianov. "4.4-μm Raman laser based on hollow-core silica fibre", Quantum Electronics **47** (5), 491-494 (2017)